\documentclass[conference]{IEEEtran}
\IEEEoverridecommandlockouts
\usepackage{cite}
\usepackage{amsmath,amssymb,amsfonts}
\usepackage{algorithmic}
\usepackage{graphicx}
\usepackage{textcomp}
\usepackage[table]{xcolor}
\usepackage{pifont}
\usepackage{array}
\usepackage{threeparttable}
\usepackage{subcaption}
\usepackage{float}

\newcommand{\XSolidBrush}{\ding{55}} 
\def\BibTeX{{\rm B\kern-.05em{\sc i\kern-.025em b}\kern-.08em
    T\kern-.1667em\lower.7ex\hbox{E}\kern-.125emX}}
\definecolor{lightgray}{gray}{0.8}

\begin{document}

\title{An Analog and Digital Hybrid \\
Attention Accelerator for Transformers \\
with Charge-based In-memory Computing
}

\author{\IEEEauthorblockN{Ashkan Moradifirouzabadi}
\IEEEauthorblockA{\textit{Electrical and Computer Engineering} \\
\textit{University of California, San Diego}\\
La Jolla, CA, USA \\
ashkan@ucsd.edu}
\and
\IEEEauthorblockN{Divya Sri Dodla}
\IEEEauthorblockA{\textit{Electrical and Computer Engineering} \\
\textit{University of California, San Diego}\\
La Jolla, CA, USA \\
ddodla@ucsd.edu}
\and
\IEEEauthorblockN{Mingu Kang}
\IEEEauthorblockA{\textit{Electrical and Computer Engineering} \\
\textit{University of California, San Diego}\\
La Jolla, CA, USA \\
mingu@ucsd.edu}
}

\IEEEaftertitletext{\vspace{-30pt}}

\makeatletter
\def\ps@IEEEtitlepagestyle{%
  \def\@oddfoot{\mycopyrightnotice}%
  \def\@oddhead{\hbox{}\@IEEEheaderstyle\leftmark\hfil\thepage}\relax
  \def\@evenhead{\@IEEEheaderstyle\thepage\hfil\leftmark\hbox{}}\relax
  \def\@evenfoot{}%
}
\def\mycopyrightnotice{%
  \begin{minipage}{\textwidth}
  \centering \scriptsize
  Copyright~\copyright~2024 IEEE. Personal use of this material is permitted. Permission from IEEE must be obtained for all other uses, in any current or future media, including\\reprinting/republishing this material for advertising or promotional purposes, creating new collective works, for resale or redistribution to servers or lists, or reuse of any copyrighted component of this work in other works by sending a request to pubs-permissions@ieee.org.
  \end{minipage}
}
\makeatother

\maketitle

\begin{abstract}
The attention mechanism is a key computing kernel of Transformers, calculating pairwise correlations across the entire input sequence. 
The computing complexity and frequent memory access in computing self-attention put a huge burden on the system especially when the sequence length increases.
This paper presents an analog and digital hybrid processor to accelerate the attention mechanism for transformers in 65nm CMOS technology.
We propose an analog computing-in-memory (CIM) core, which prunes ~75\% of low-score tokens on average during runtime at ultra-low power and delay.
Additionally, a digital processor performs precise computations only for ~25\% unpruned tokens selected by the analog CIM core, preventing accuracy degradation.
Measured results show peak energy efficiency of 14.8 and 1.65 TOPS/W, and peak area efficiency of 976.6 and 79.4 GOPS/mm$^\mathrm{2}$ in the analog core and the system-on-chip (SoC), respectively.
\end{abstract}


\vspace{-5pt}
\section{Introduction}
\vspace{-5pt}

Transformer models excel in various Natural Language Processing (NLP) and Computer Vision (CV) tasks, achieving top performance due to their attention mechanism. 
This mechanism processes the input data by converting it into three distinct embedding matrices: query ($Q$), key ($K$), and value ($V$). 
The fundamental operation of the attention mechanism involves assessing the correlation between the query and key. Following this evaluation, it computes a weighted sum of the value vectors to incorporate the determined correlations.
%\cite{vaswani2017attention}.

Despite their success, the high computational and memory demands of attention, which increase quadratically with sequence length, pose significant challenges to hardware efficiency.
Recent studies suggest that the majority of tokens can be pruned during runtime by evaluating the attention score ($S$), which is determined by the product of the query and key embeddings ($Q.K^T$) \cite{a3:hpca20,spatten:hpca21,sundaram2023freflex,yazdanbakhsh2022sparse}.
Nevertheless, these are feasible only 
 after fetching the vectors from the memory and performing dot-product, limiting the practical efficacy.

%none of these studies explore the benefits of conducting such pruning directly within the on-chip memory by implementing a custom SRAM design.

\begin{figure}[t]
\centering
\includegraphics[width=0.85\columnwidth]{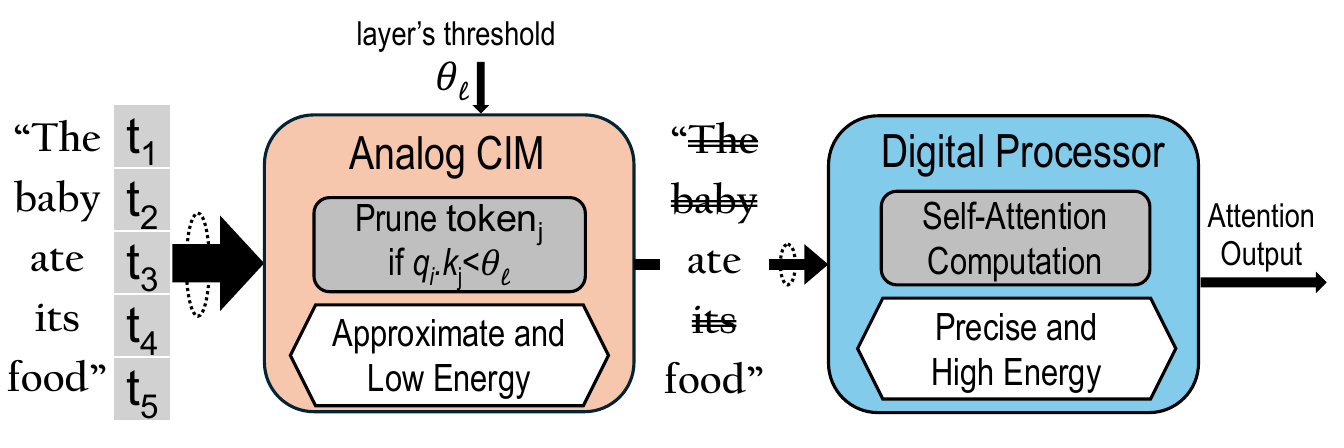}
\caption{Overview of the token pruning mechanism using analog in-memory computing and hybrid digital processing.}
\label{fig:overal} 
\vspace{-15pt}
\end{figure}

We introduce an analog computing-in-memory (CIM) core for early detection of weakly-correlated tokens within the SRAM memory bank, as shown in Fig. \ref{fig:overal}, operating with high parallelism and consuming less than 8\% of the total power.
Assuming $q_i$ and $k_j$ are two column vectors from $Q$ and $K$ matrices, the CIM core decides whether to prune $k_j$ for $q_i$ based on their dot-product value.

Analog CIM often faces hurdles such as high analog-to-digital converter (ADC) costs and inaccuracies due to the non-idealities from analog processing.
Our design overcomes these challenges by performing binary pruning decisions through an analog comparator, thereby eliminating the need for expensive ADCs.
Additionally, a high-precision digital processor calculates the remaining high-score tokens to retain accuracy.
Also, we enhance the SNR and accuracy of analog CIM pruning by the sparsity-aware selective charge sharing (SSCS) technique, ensuring no accuracy degradation occurs in the NLP task.

\vspace{-5pt}
\section{Proposed Design}
\vspace{-5pt}
\subsection{Microarchitecture}
\vspace{-3pt}
Fig. 2 outlines the design and data flow of the proposed hybrid processor. 
On the analog side, each query vector $q_i$ is sequentially transferred from the $Q$-buffer to the CIM core for in-memory computation with all key tokens in the array. In this process, only the 4 most significant bits (MSB) out of 8 bits are stored and utilized in the CIM core for both $q_i$ and $k_j$ vectors.
The bitline processor (BLP) then calculates a binary-weighted-sum (BWS) of the CIM array's analog output, taking into account the bit position of each RBL. The output of BLP indicates the analog representation of the attention score for $q_i$ with all other tokens.
Finally, a comparator evaluates if the attention score of each token ($q_i$.$k_j$) exceeds a specific threshold|a value derived from model training|to decide on pruning the tokens with low score magnitudes.
The outcome of the comparators is a 64-bit binary vector $U$, where $u_j = 1$ indicates that the corresponding $k_j$ token is not pruned.

In the digital core, the data overlap detection engine monitors which tokens are already present in the local register. A $k_j$ vector is retrieved from the CIM core and sent to the digital processor only if indicated by $U$ and it does not reside in the register at the moment.
Notably, over 80\% of unpruned tokens are found to be common across consecutive queries, which significantly minimizes the requirement for fetching new data.
The high-precision digital processor then recalculates the product of the $q_i$ vector with the $K$ matrix, only for the unpruned tokens. The process is done using 8-bit $q$ and $k$ values by incorporating the least significant 4 bits of $k$ from a standard SRAM bank.
The subsequent processing stages including applying the \textit{Softmax} function and multiplying with value embeddings ($V$) are also performed in the digital processor only for the unpruned tokens, all in a pipeline.

It is important to note that the proposed design supports simultaneous operations in both CIM and standard read modes. As a result, while unpruned key embeddings for $q_i$ are being fetched from the CIM core through a standard read operation, the CIM core is concurrently processing the attention scores for the next query, $q_{(i+1)}$.

\begin{figure}[t]
\centering
\includegraphics[width=0.85\columnwidth]{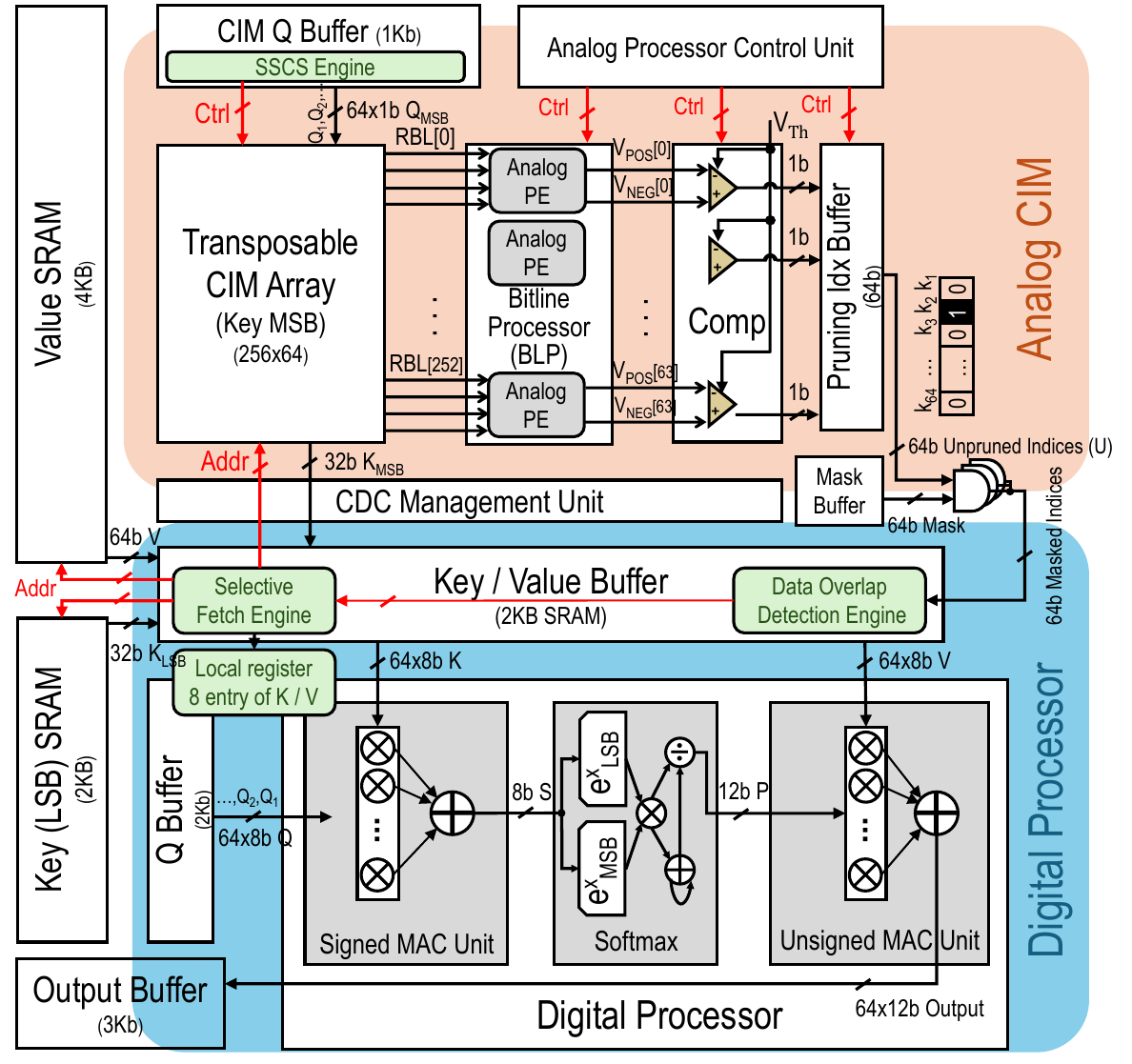}
\caption{Overall architecture and dataflow of the design.}
\label{fig:arch} 
\vspace{-15pt}
\end{figure}

\vspace{-5pt}
\subsection{Transposable Memory Array}
\begin{figure}[t]
\centering
\includegraphics[width=0.95\columnwidth]{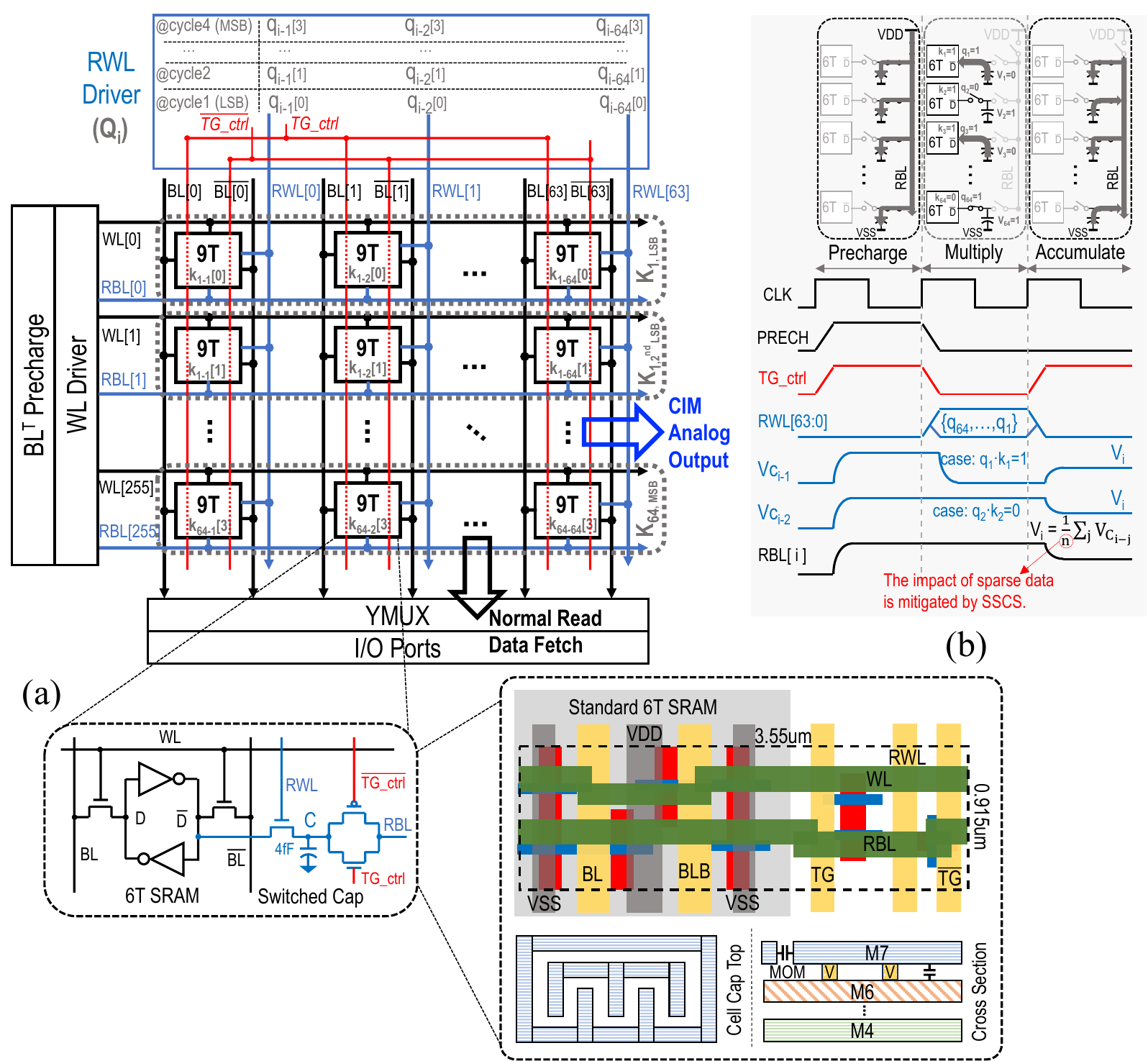}
\caption{Transposable memory array. (a) The array architecture supporting CIM and standard read operations with the proposed 9-T bitcell, (b) The timing diagram of CIM operation.}
\label{fig:cim} 
\vspace{-15pt}
\end{figure}

\vspace{-2pt}
Fig. \ref{fig:cim} shows the memory array with our proposed 9-T bitcell that supports both analog in-memory vector-matrix multiplication and standard read/write operations.
Each 4-bit $k_j$ vector with 64 elements is stored across four rows and 64 columns in the array. 
After determining which tokens should not be pruned by CIM, the digital processor initiates a standard read operation to acquire the pertinent unpruned $k_j$ vectors. 
These vectors are organized within four neighboring rows in the memory array and are accessed through vertical routing along the bitlines (BLs).
Conversely, CIM operations extract the analog output using horizontal routing. To accommodate the different directions required for data retrieval between standard read and CIM operations, the bitcell array is equipped with wordlines (WLs) and bitlines (BLs) for normal reading and writing while the CIM operations use read wordlines (RWLs) and read bitlines (RBLs), which are oriented perpendicularly to each other.

The CIM operation consists of three main phases: precharge, multiply, and accumulate.
After precharging the bitcell capacitors, analog multiplication is executed by broadcasting the binary $q$ values ($q_{i-n}[b]$\footnote{$q_i$ vector's $n^{th}$ element's $b^{th}$ bit. $k$ also has the same notation.}) sequentially via RWL driver, starting with LSB.
Capacitors discharge only when both stored $k$ data ($k_{j-n}[b]$) and the broadcasted $q$ are high, effectively storing the inverse of the binary product ($qk$).
It is important to note that the SRAM is sized to ensure no data loss when RWL is high in any condition.
In the accumulation phase, capacitors of the bitcells in the same row engage in charge-sharing when $TG\_ctrl$ is activated. This process establishes the binary dot-product by aggregating the charge across all columns connected to an RBL, so that a voltage drop from the initially precharged level on the RBL indicates the computed result. Following this phase, the array progresses to the next bit position of $q_i$ in the following cycle, processing bits in order from LSB to MSB.

\vspace{-5pt}
\subsection{Bitline Processor}
\vspace{-3pt}
Fig. 4 illustrates the operation of the binary-weighted sampler (BWS) in BLP to complete the dot-product with 4-bit elements in $q_i$ and $k_j$ vectors.
The bit positions of $k_j$ are mapped row-wise across four consecutive RBLs, while $q_i$'s bit positions are cycled through.
The BWS uses equal capacitance for sampling ($C_{SP}$) and storage ($C_{ST}$), and operates in three phases: refresh, sample, and store. $C_{SP}$ is refreshed before each sample phase to clear any residual charge.
Then, the first input voltage ($V_{in}[0]$) is sampled on $C_{SP}$. During the store phase, $V_{in}[0]$ from $C_{SP}$ is shared with $C_{ST}$, producing an output of $0.5V_{in}[0]$.
After another sampling phase, the newly sampled $V_{in}[1]$ on $C_{SP}$ is shared with the stored voltage on $C_{ST}$, generating an output of $0.5^2V_{in}[0]+0.5V_{in}[1]$.
This process repeats for each analog input voltage, leading to an output representing the binary-weighted sum of the four sequentially provided analog inputs: $0.5^4V_{in}[0]+0.5^3V_{in}[1]+0.5^2V_{in}[2]+0.5V_{in}[3]$.

\begin{figure}[t]
\centering
\includegraphics[width=0.85\columnwidth]{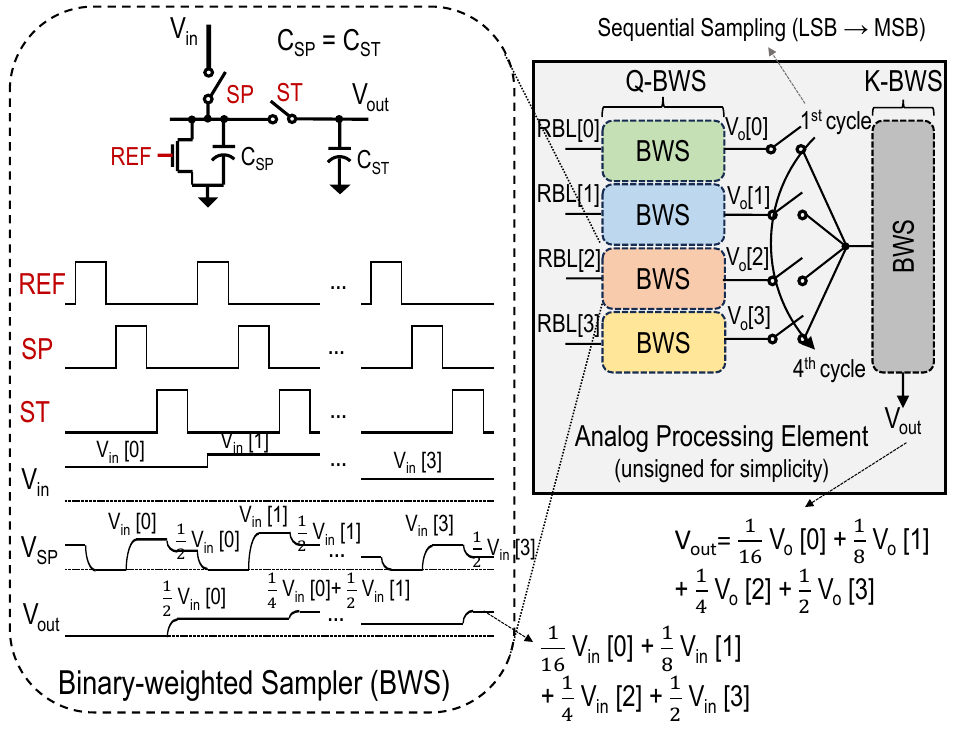}
\caption{Bitline processor (BLP) for binary-weighted summation with a timing diagram.}
\label{fig:blp} 
\vspace{-15pt}
\end{figure}

Each BWS in $Q$-BWS computes the binary-weighted sum as above for the sequentially provided RBL outputs to represent $q_i$'s bit positions.
Following the completion of sampling across all RBLs, another BWS mechanism ($K$-BWS, depicted in gray in Fig. \ref{fig:blp}) establishes the binary weights across the outputs of four $Q$-BWS. This process starts by sampling the LSB $V_o[0]$ and ends with the MSB $V_o[3]$, thus representing the bit positions of $k_j$. The $K$-BWS reuses the $C_{ST}$ from the $Q$-BWS as its sampling capacitors ($C_{SP}$), avoiding the need for separate $C_{SP}$ within its structure.

Since $q_i$ and $k_j$ include signed 4-bit elements with the MSB weighted as ‘-8’, the cycle and RBL position determine whether the RBL output signifies a positive or negative value.
Therefore, in practice, a signed $Q$-BWS is implemented by using two $C_{ST}$ to store these positive and negative values separately. These values are then processed through two separate sets of $K$-BWS to represent the bit positions of $k_j$, as described earlier. 
Following this, the positive and negative outputs are applied to the comparator to make binary pruning decisions based on the threshold voltage.

\vspace{-5pt}
\section{Measurement Results}
\vspace{-5pt}

\begin{figure}[t]
\centering
\includegraphics[width=0.99\columnwidth]{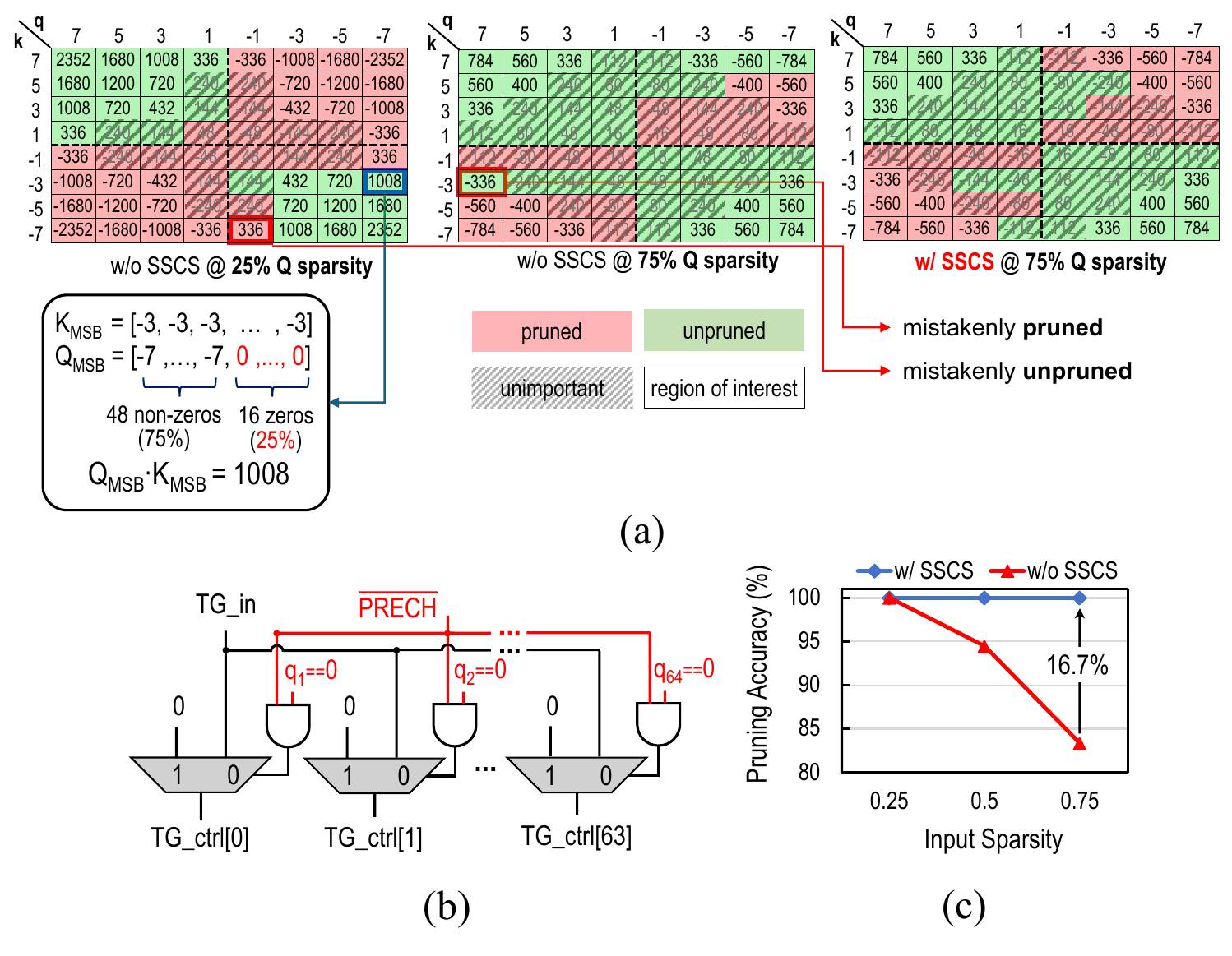}
\caption{Pruning accuracy of the CIM core. (a) Pruning decision map of the comparator with different $q$ and $k$, and a threshold of 0. (b) SSCS circuitry. (c) The effect of incorporating SSCS on the pruning accuracy with different input sparsities.}
\label{fig:prun} 
\vspace{-15pt}
\end{figure}

The accelerator is fabricated in 65nm CMOS technology. 
Fig. \ref{fig:prun} shows the measured pruning decisions of the comparator assuming a pruning threshold of zero. The region of interest is based on the required decision resolution. 
The software simulation indicates that a 9-bit resolution for $Q$.$K^T$ (out of 14-bit output) is sufficient for the pruning purpose to maintain the application's final accuracy. As a result, we ignore the values between -256 and 255 as misidentification of them does not affect the accuracy.
It should be noted that a higher sparsity in $q$ leads to smaller output values (near-zero output voltage), which degrades the comparator's accuracy. 
To counteract this, the SSCS technique is employed to exclude zero-magnitude $q$s during charge-sharing in the CIM operation (accumulate phase). This approach enhances pruning accuracy by 15.6\% in our measurements, achieving a 0\% pruning error rate for the target 9-bit resolution. 

\begin{figure}[t]
\centering
\includegraphics[width=0.45\columnwidth]{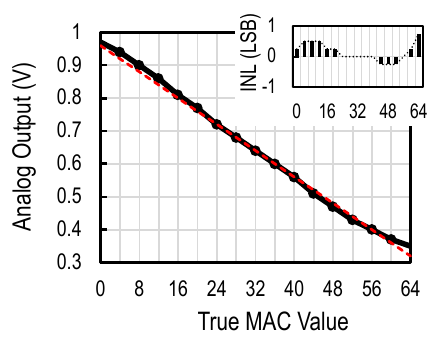}
\caption{The analog output of RBL compared to the expected MAC values (red dotted line).}
\label{fig:rbl} 
\vspace{-10pt}
\end{figure}

\begin{table}[h]
\centering
\caption{Measured Application Accuracy (Baseline: INT8).}
\vspace{-5pt}
\label{tab:accuracy}
\begin{tabular}{|m{2.7cm}|m{1.5cm}|m{1.5cm}|m{1.5cm}|}
%\cline{1-4}
\hline
\centering\cellcolor{lightgray} NLP Model & \multicolumn{3}{m{4.5cm}|}{\centering BERT-Base}
\tabularnewline
%\cline{2-4}
\hline
\centering\cellcolor{lightgray} NLP Benchmark & \multicolumn{3}{m{4.5cm}|}{\centering General Language Understanding Evaluation (GLUE)}
\tabularnewline
\hline
\centering\cellcolor{lightgray}NLP Task & \centering CoLA Grammatical Correction & \centering MRPC Paraphrase Analysis & \centering SST-2 Sentiment Analysis
\tabularnewline
\hline
\centering\cellcolor{lightgray}Baseline Accuracy (\%) & \centering 81.86\% & \centering 85.56\% & \centering 92.29\%
\tabularnewline
\hline
\centering\cellcolor{lightgray}Proposed Accuracy\\w/ CIM Pruning (\%) & \centering 81.48\% & \centering 85.05\% & \centering 91.96\%
\tabularnewline
\hline
\centering\cellcolor{lightgray}Pruning Rate & \centering 77.23\% & \centering 70.06\% & \centering 81.28\%
\tabularnewline
\hline
\end{tabular}
%\vspace{-15pt}
\end{table}

\begin{table*}[ht]
\centering
\begin{threeparttable}
\caption{Comparison with Prior Works.}
\vspace{-15pt}
\label{tab:comparison}
\begin{tabular}{|m{3.5cm}|m{1.6cm}|m{1.6cm}|m{1.6cm}|m{1.6cm}|m{2.7cm}|}
\hline
\centering\cellcolor{lightgray} & \centering\cellcolor{lightgray}VLSI'22\cite{wang202232} & \centering\cellcolor{lightgray}ISSCC'22\cite{wang202228nm} &
\centering\cellcolor{lightgray}ISSCC'23\cite{tambe202322} & 
\centering\cellcolor{lightgray}JSSC'23\cite{sundaram2023freflex} & 
\centering\cellcolor{lightgray}\textbf{This Work} \tabularnewline
\hline
\centering\cellcolor{lightgray}Technology & \centering 22nm & \centering 28nm & \centering 12nm & \centering 65nm & \centering\textbf{65nm} \tabularnewline
\hline
\centering\cellcolor{lightgray}Application & \centering General & \centering Transformer & \centering Transformer & \centering Transformer, CNN & \centering\textbf{Transformer} \tabularnewline
\hline
\centering\cellcolor{lightgray}Runtime Pruning & \centering\textbf{-} & \centering\XSolidBrush & \centering\XSolidBrush & \centering Digital Token Pruning & \centering\textbf{Analog In-memory Token Pruning} \tabularnewline
\hline
\centering\cellcolor{lightgray}Implementation & \centering Charge-based Analog CIM & \centering Digital & \centering Digital & \centering Digital & \centering\textbf{Charge-based Analog CIM + Digital}  \tabularnewline
\hline
\centering\cellcolor{lightgray}Precision [Input:Weight] & \centering Analog[8:8] & \centering INT12 & \centering FP4/FP8 & \centering INT8 & \centering\textbf{Analog[4:4], INT8} \tabularnewline
\hline
\centering\cellcolor{lightgray}Die Area [mm$^\mathrm{2}$] & \centering 0.25 & \centering 6.82 & \centering 4.60 & \centering 6.40 & \centering\textbf{3.20} \tabularnewline
\hline
\centering\cellcolor{lightgray}Supply Voltage [V] & \centering 0.7 - 1.1 & \centering 0.56 - 1.1 & \centering 0.62 - 1.0 & \centering 0.9 - 1.2 & \centering\textbf{0.7 - 1.2} \tabularnewline
\hline
\centering\cellcolor{lightgray}Frequency [MHz] & \centering 145 - 240 & \centering 50 - 510 & \centering 77 - 717 & \centering 600 - 1100 & \centering\textbf{350 - 1100} \tabularnewline
\hline
\centering\cellcolor{lightgray}Power [mW] & \centering \textbf{-} & \centering 12 - 273 & \centering 9 - 111 & \centering 400 - 1675 & \centering\textbf{49 - 456} \tabularnewline
\hline
%\cellcolor{lightgray}Performance (TOPS) & \centering 1.0\\(0.49\tnote{*}) & \centering 0.522\\(0.27$^*$) & \centering 0.734\\(0.16$^*$) & \centering 1.02 & \centering\textbf{0.25} \tabularnewline
%\hline
\centering\cellcolor{lightgray}Energy Efficiency [TOPS/W] & \centering CIM: 32.2\\(11.4$^*$)$^1$ & \centering 4.25 - 27.56$^2$\\(0.5 - 3.23)$^*$ & \centering 18.1\\(0.32$^*$) & \centering 0.6 - 1.0 & \centering\textbf{CIM: 14.8\\SoC: 1.65} \tabularnewline
\hline
\centering\cellcolor{lightgray}Area Efficiency [GOPS/mm$^\mathrm{2}$] & \centering CIM: 4000\\(278.5$^*$) & \centering 76.5$^3$\\(3.9$^*$) & \centering 159.57$^3$\\(2.58$^*$) & \centering 160 & \centering\textbf{CIM: 976.6\\SoC: 79.4} \tabularnewline
\hline
\end{tabular}
\begin{tablenotes}
\footnotesize{
\item $^*$Scaled to 65nm by the rules in “http://vcl.ece.ucdavis.edu/pubs/2017.02.VLSIintegration.TechScale/” -- $^1$Scaled to [4:4] for fair comparison -- $^2$Matrix multiplication with 90\% sparsity -- $^3$Calculated based on the provided performance and area values
}
\end{tablenotes}
\end{threeparttable}
\vspace{-15pt}
\end{table*}

\begin{figure}[t]
\centering
\includegraphics[width=0.85\columnwidth]{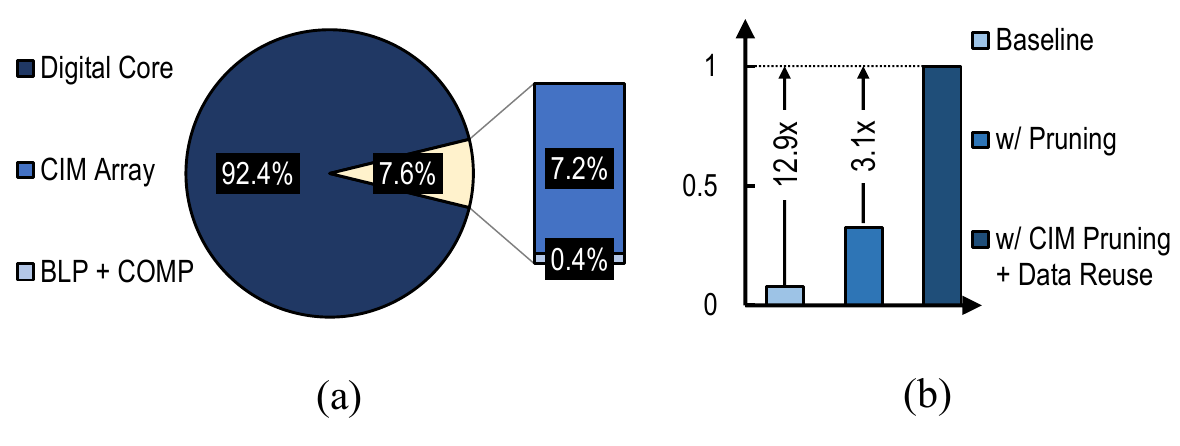}
\caption{Measured behavior of the chip at the nominal condition (@1V and 840MHz). (a) Power breakdown, (b) Normalized energy efficiency.}
\label{fig:power} 
\vspace{-15pt}
\end{figure}

\begin{figure}[t]
\centering
\includegraphics[width=0.55\columnwidth]{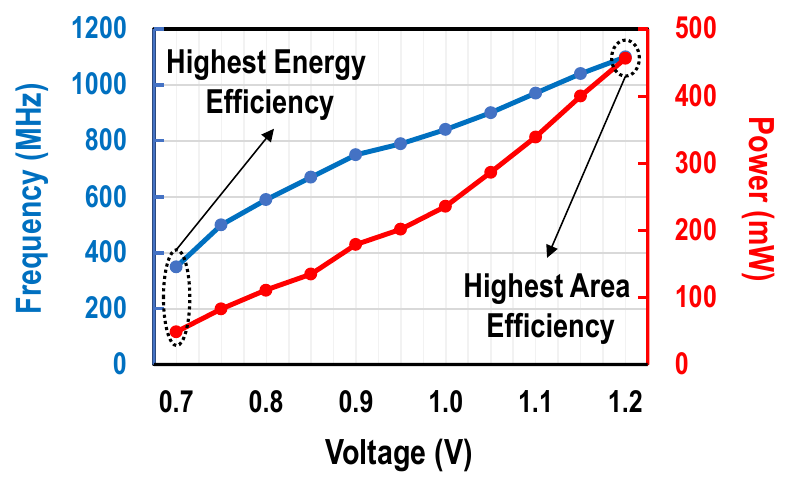}
\caption{Voltage-frequency scaling and power consumption of each operational point.}
\label{fig:vfs} 
\vspace{-15pt}
\end{figure}

\vspace{-2pt} 
Fig. \ref{fig:rbl} displays the RBL output from the CIM operation, which demonstrates satisfactory linearity for the target resolution.
Table \ref{tab:accuracy} shows the accuracy on three datasets, with a negligible accuracy loss ($<$0.38\%) and 70.1-81.3\% pruning rates. 
Fig. \ref{fig:power} presents 12.9× and 3.1× energy savings over 8-b fully digital hardware implementation without and with pruning, respectively. These energy benefits arise from analog CIM pruning and unpruned token reuse, with the CIM core adding only 7.6\% power overhead. 
Fig. \ref{fig:vfs} shows the measured voltage-frequency scaling (VFS) and power consumption at different voltages.

Table \ref{tab:comparison}, compares our work with the previous chips targeting Transformers \cite{wang202228nm,tambe202322,sundaram2023freflex}.
We also include a charge-based analog in-memory computing design \cite{wang202232} to evaluate the performance of our analog core.
Our design is the first to use charge-based analog CIM in SRAM and hybrid digital implementation for Transformer application.
Although the previous works mostly use more advanced technology nodes, our design outperforms most of them in energy and area efficiency considering the scaling rule for a fair comparison, delivering 1.65 and 14.8 TOPS/W, and 79.4 and 976.6 GOPS/mm$^\mathrm{2}$ for the system and CIM respectively, with the maximum practical frequency of 1.1GHz.

\begin{figure}[t]
\centering
\includegraphics[width=0.75\columnwidth]{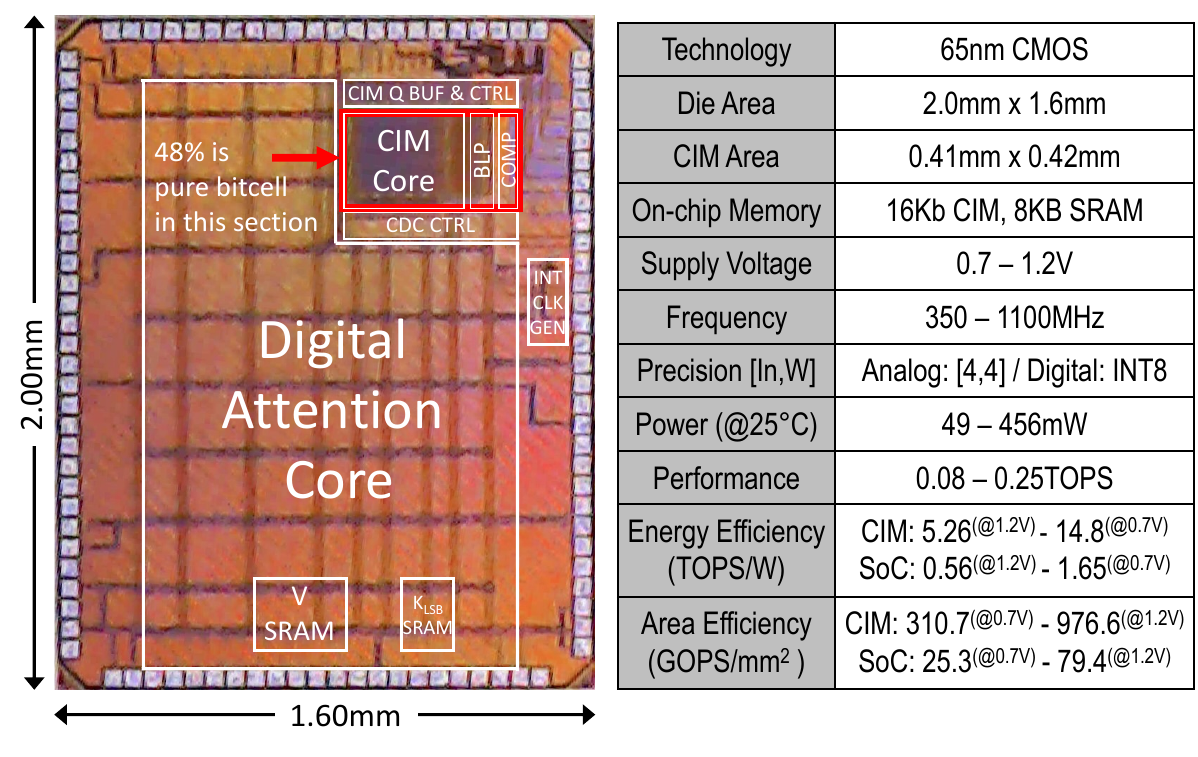}
\vspace{-5pt}
\caption{Die photo and chip specifications.}
\label{fig:die} 
\vspace{-15pt}
\end{figure}
\vspace{-10pt}
\section{Conclusion}
\vspace{-5pt}

This paper proposes a hybrid analog and digital processor to accelerate the attention mechanism in Transformers.
By using a custom 9-T SRAM cell that enables charge-based analog in-memory computing, we can skip around 75\% of the weakly-related tokens in the low-power and highly parallel analog core. At the same time, a digital high-precision processor performs the attention computation only for the ~25\% important tokens, maintaining on-par accuracy.

\vspace{-10pt}
\section{Acknowledgement}
\vspace{-4pt}

This work was supported by the PRISM Center within JUMP 2.0, a Semiconductor Research Corporation (SRC) program in cooperation with DARPA, and by the National Research Foundation of Korea (NRF) grant funded by the Korean government (MSIT) (RS-2024-00347090).
\vspace{-5pt}

\bibliographystyle{IEEEtranS}
{\footnotesize
\bibliography{refs}}

\end{document}